\documentclass[aps,pra,twocolumn,superscriptaddress,showpacs]{revtex4-2}

\usepackage[utf8]{inputenc}

\usepackage{array}
\usepackage{amsmath}
\usepackage{graphicx}
\usepackage{epstopdf}
\usepackage{color}
\usepackage{amsfonts}
\usepackage{subfigure}
\usepackage{amssymb}
\usepackage{graphicx}
\usepackage{makecell}
\usepackage[colorlinks,citecolor=blue]{hyperref}
\usepackage{changes}
\def\ket#1{\mathinner{|{#1}\rangle}}

\newcommand{\onecolm}{
  \end{multicols}
  \vspace{-3.5ex}
  \noindent\rule{0.5\textwidth}{0.1ex}\rule{0.1ex}{2ex}\hfill
}
\newcommand{\twocolm}{
  \hfill\raisebox{-1.9ex}{\rule{0.1ex}{2ex}}\rule{0.5\textwidth}{0.1ex}
  \vspace{-4ex}
  \begin{multicols}{2}
}

\begin{document}

\title{Tunable chiral spiral phases in a non-Hermitian Ising-Gamma spin chain}
\date{\today}

\author{Run-Dong Huang}

\affiliation {Key Laboratory of Atomic and Subatomic Structure and Quantum Control (Ministry of Education), Guangdong Basic Research Center of Excellence for Structure and Fundamental Interactions of Matter, School of Physics, South China Normal University, Guangzhou 510006, China}
\affiliation {Guangdong Provincial Key Laboratory of Quantum Engineering and Quantum Materials, Guangdong-Hong Kong Joint Laboratory of Quantum Matter, Frontier Research Institute for Physics, South China Normal University, Guangzhou 510006, China}

\author{Wei-Lin Li}

\affiliation {Key Laboratory of Atomic and Subatomic Structure and Quantum Control (Ministry of Education), Guangdong Basic Research Center of Excellence for Structure and Fundamental Interactions of Matter, School of Physics, South China Normal University, Guangzhou 510006, China}
\affiliation {Guangdong Provincial Key Laboratory of Quantum Engineering and Quantum Materials, Guangdong-Hong Kong Joint Laboratory of Quantum Matter, Frontier Research Institute for Physics, South China Normal University, Guangzhou 510006, China}

\author{Zhi Li}
\email{lizphys@m.scnu.edu.cn}
\affiliation {Key Laboratory of Atomic and Subatomic Structure and Quantum Control (Ministry of Education), Guangdong Basic Research Center of Excellence for Structure and Fundamental Interactions of Matter, School of Physics, South China Normal University, Guangzhou 510006, China}
\affiliation {Guangdong Provincial Key Laboratory of Quantum Engineering and Quantum Materials, Guangdong-Hong Kong Joint Laboratory of Quantum Matter, Frontier Research Institute for Physics, South China Normal University, Guangzhou 510006, China}

\begin{abstract}
We study the influence of dissipation on the Ising-Gamma model. Through observables such as ground-state energy, order parameters, entanglement entropy, etc., we identify each phase region and provide the global phase diagram of the system. The results show that the region of the spiral phase will continuously expand with the increase of dissipation, gradually squeezing the original paramagnetic and antiferromagnetic phase regions. Remarkably, unlike the conservative system, the introduction of dissipation will cause two spiral phases with opposite chirality to emerge simultaneously in the system, which provides a possibility for the manipulation of spiral chirality in cold atomic experiments. Moreover, we reveal the dependence mechanism of the transformation between these two spiral phases with distinct chirality on the strength of the relative coefficient of off-diagonal Gamma interactions in the Ising-Gamma model. Since both the relevant order parameters and dissipation can be well controlled within a detectable range, these phenomena can be observed in ultracold atomic experiments.
\end{abstract}

\maketitle


\section{INTRODUCTION}
The investigation of exotic quantum phases and critical phenomena constitutes one of the pivotal research directions in modern condensed matter physics and statistical mechanics. Serving as a fundamental model for describing phase transitions and magnetic systems, the Ising model has been extensively employed in relevant theoretical studies. Notably, to extend the understanding of non-trivial excitations and topological order in strongly-correlated quantum systems, the research focus has progressively expanded to generalized models that incorporate more complex spin interactions. Among these, the off-diagonal Gamma exchange interactions have attracted significant attention in recent years~\cite{LLL2011,YWT2020,LYY2020,SAK2021,LZK2021,LYY2021,Zhaozhuan2022,KHN2024,SHC2024,MSM2025}. These types of bond-dependent interaction involve the cross-coupling between different spin components on neighboring lattice sites and was initially systematically proposed in the Kitaev honeycomb model~\cite{GG2009,HMJ2018}. The integration of such interactions with Ising-type interactions constitutes the Ising-Gamma model, which serves as an ideal platform for studying quantum many-body effects that go beyond traditional magnetic order. Studies have shown that the off-diagonal Gamma interaction can induce a variety of novel quantum phases and phase transition behaviors, such as stabilizing quantum spin liquids~\cite{GG2009,LJX2021,TDS2019,WYL2021}, giving rise to magnetic long-range order~\cite{WYL2021,RKC2006,HKS2011} and inducing gapless chiral phases~\cite{FOA2012,LQ2022}. Furthermore, such interactions not only hold the potential for theoretical tractability but can also be realized in experimental platforms such as trapped atoms~\cite{Zhaozhuan2022,KHY2024}, photonic lattice systems~\cite{PLA2017,DJH2015,GAC2015}, and solid-state material systems including specific transition metal compounds~\cite{KMA2023,YHQ2023,DVE2014,BDG1996,CRD2010,KMR2002,BOM2013}, thus offering a feasible pathway for exploring novel quantum states of matter and critical phenomena from both theoretical and experimental perspectives.

On the other hand, dissipation is unavoidable in almost all experimental platforms, including ultracold atomic systems~\cite{Li2019,LLC2020}, optical systems~\cite{Chen2017,PBF2014,CAS2019,Zhen2015}, nitrogen-vacancy center~\cite{WYW2019}, and trapped ions~\cite{LLY2022,DCC2022,Zhang2022}. Therefore, non-Hermiticity can no longer be regarded as a negligible component in theoretical models but has emerged as one of the fundamental aspects in the exploration of novel quantum phenomena. Recently, a large number of studies have focused on non-Hermitian quantum systems, not only because they are highly relevant to experimental reality, but also because they can present a series of unique physical phenomena that do not exist in the traditional Hermitian framework. These physical phenomena include: boundary-localized accumulation of bulk states driven by the non-Hermitian skin effect~\cite{YJY2025,QL2022,Longhi2024,Ma2024,Zhang2023}, new topological structures closely related to exception points~\cite{Chen2017,Li2023,MAM2019,KKK2019,ZLZ2020,He2020,SHB2018}, disorder phenomena generated by non-Hermitian conditions~\cite{Luo2019,JHZ2019,Lee2014,LGJ2024} and mobility edge induced by non-Hermiticity~\cite{lsz2404,LGJ2024,PLL2025,ZLL2025}. Notably, the influence of non-Hermiticity on quantum phase transition processes has progressively emerged as a prominent research focus in recent years. For instance, in spin models incorporating complex fields, non-Hermitian perturbations can substantially alter the critical properties of the system, even inducing novel critical points or topological quantum phase transitions~\cite{GYL2022,LCL2025}. Therefore, the development of theoretical models that simultaneously incorporate strong correlation effects and non-Hermitian properties is of paramount importance. Such a model would not only help elucidate existing experimental observations but also pave the way for new theoretical frameworks, thereby facilitating a systematic exploration of dissipation-induced emergent quantum states and their associated critical phenomena.

So far, although both the off-diagonal Gamma interactions and non-Hermitian studies have come under the spotlight, few efforts are put to explore the properties of the systems by combining the two \cite{KADS2023,KADS2025,AAA2024}. This work is devoted to the ground-state properties and phase transitions of the dissipative Ising-Gamma model.  We will construct a non-Hermitian Ising-Gamma model by introducing a complex field. Then, by calculating the energy gap, we present the non-Hermitian phase diagram.


The rest of this manuscript is organized as follows. Sec.~\ref{sec:model} introduces the model and studies its ground-state properties through analytical calculations. In Sec.~\ref{sec:observables}, the introduction of observables is provided. Then, we overview the phase diagram in Sec.~\ref{sec:phase}. We calculate order parameters and entanglement entropy in Sec.~\ref{sec:order} to identify the properties of different phases. In Sec.~\ref{sec:variation}, we investigate the phase transitions and critical behaviors. Sec.~\ref{sec:summary} is the summary of this paper. 


\section{MODEL AND analytical solution}
\label{sec:model}
In the following, we start with a dissipative Ising-Gamma model (see Appendix~\ref{sec:app0} for details) with exactly solvable ground-states, whose Hamiltonian reads:
\begin{equation}\label{Hami}
\begin{aligned}
H_{\rm eff}=& \sum_{j=1}^N J \sigma^x_j\sigma^x_{j+1}+\sum_{j=1}^N\Gamma\left(\sigma^x_j\sigma^y_{j+1}+\alpha\sigma^y_j\sigma^x_{j+1}\right)\\
&+\sum_{j=1}^N\left(h\sigma^z_j-\frac{i\eta}{2}\sigma^u_j\right),
\end{aligned}
\end{equation}
where $\sigma^{x}_{j}$, $\sigma^{y}_{j}$ and $\sigma^{z}_{j}$ are the Pauli matrices of the $j$th spin.
$\sigma^{u}_j=\begin{bmatrix}
1 & 0  \\
0 & 0
\end{bmatrix}$ denotes loss or gain effect, which can be conveniently realized in optical systems and optical lattice ultracold atomic systems~\cite{Luo2019,P2012,Z2015}. 
Without loss of generality, we take $J=1$ as the unit of energy in the following calculation. We set the amplitude of off-diagonal Gamma interactions $\Gamma=1$ for simplicity. Experimentally, there are three controllable parameters, namely, the relative coefficient of off-diagonal exchange couplings $\alpha$, the strength of the uniform transverse field $h$ and the dissipation strength $\eta$.

One can transform Eq.~\eqref{Hami} into fermionic representation by conducting a Jordan-Wigner transformation~\cite{JW1928,SL2010}, which is defined as
\begin{equation}
\begin{aligned}
    \label{jw}
    \sigma^x_j&=\prod_{l<j}\left(1-2c^{\dagger}_{l} c_{l}\right)\left(c^{\dagger}_j+c_j\right),\\
    \sigma^y_j&=i\prod_{l<j}\left(1-2c^{\dagger}_{l} c_{l}\right)\left(c^{\dagger}_j-c_j\right),\\
    \sigma^z_j&=1-2c^{\dagger}_{j} c_{j}, 
\end{aligned}
\end{equation}
where $c_j^{\dagger}$($c_j$) is the creation (annihilation) operator at site $j$. Then, one can perform Fourier transform
\begin{equation}
    c^{\dagger}_j=\frac{e^{i\pi/4}}{\sqrt{N}}\sum_k e^{-ikj}c^{\dagger}_k,
\end{equation}
then, we obtain
\begin{equation}
\begin{aligned}
 H=\sum_{k} 
\begin{pmatrix}
c_k^\dagger & c_{-k}
\end{pmatrix}
\begin{pmatrix}
I_k+z_k & x_k-iy_k \\
x_k+iy_k & I_k-z_k
\end{pmatrix}
\begin{pmatrix}
c_k \\
c_{-k}^\dagger
\end{pmatrix},
\end{aligned}
\end{equation}
where $I_k=\Gamma(1-\alpha)\sin k$, $x_k=-J\sin k$, $y_k=\Gamma(1+\alpha) \sin k$ and $z_k=J\cos k+\frac{i\eta}{4}-h$. By using Bogoliubov transformation,
\begin{equation}
    b_k=u_kc_k+v_kc^{\dagger}_{-k},~\bar{b}_{k}=u_kc^{\dagger}_k-v_kc_{-k}.
\end{equation}
Eventually, we get the diagonalized Hamiltonian~\cite{Zhaozhuan2022,GYL2022,LCL2025}
\begin{equation}
    H=\sum_{k} \Lambda_{k}(\bar{b}_{k}b_{k}-\frac{1}{2}),
\end{equation}
where
\begin{equation}\label{lambdak}
  \Lambda_k=2I_k+2\sqrt{x_k^2+y_k^2+z_k^2}.
\end{equation}
In this work, we define the ground-state as the state with the minimum real part of $\Lambda_{k}$. The ground-state of Eq.~\eqref{Hami} is
\begin{equation}\label{state}
    \ket{G}=\frac{1}{\sqrt{M}}\prod_{k>0}[u_{k}-v_{k}c^{\dagger}_{k} c^{\dagger}_{-k}]\ket{0},
\end{equation}
where $M=\prod_{k>0}(|u_{k}|^2+|v_{k}|^2)$ is the normalization constant, $u_{k}=\frac{-x_k+iy_k}{C}$, $v_{k}=\frac{I_k+z_k-\Lambda_k}{C}$ and $C$ is a constant to satisfy $u_{k}^{2}+v_{k}^{2}=1$.

\section{OBSERVABLES AND METHODS}\label{sec:observables}
\subsection{The ground-state energy density and its second-order derivative}
According to Eq.~\eqref{lambdak}, the ground-state energy density can be defined as
\begin{equation}
e_0=-\frac{1}{2N}\sum_{k}|\Lambda_k|,
\end{equation}
and we can easily obtain the second derivative of $e_0$ with respect to $h$, i.e., $\frac{\partial^2 e_0}{\partial h^{2}}$.

\subsection{Energy gap}
For Hermitian systems, the minimum value of $\Lambda_{k}$ is defined as the energy gap, which is conventionally denoted by $\Delta$, i.e., 
\begin{equation}
\Delta=\min_{k}\Lambda_{k}.
\end{equation}
The closing of the energy gap ($\Delta=0$) typically signifies either a phase transition critical point or the emergence of a gapless phase. In non-Hermitian cases, however, since the value of $\Lambda_{k}$ is complex, the gap of the corresponding emergent phases is complex. In this paper, we only consider the real part of the energy gap.

\subsection{Order parameters}
To identify each phase, we calculate the spin correlation function and the cross-correlation, which are two key quantities to study Ising-Gamma spin model. The spin correlation function is defined as
\begin{equation}
G^{\alpha \beta}_r=\left\langle\sigma_{j}^{\alpha} \sigma_{l}^{\beta}\right\rangle,
\end{equation}
where $r=l-j$, $\alpha$, $\beta=x,~y,~z$. Then, one can obtain
\begin{equation}
\begin{aligned}
G^{xx}_r &=\left\langle\left(c_{j}^{\dagger}-c_{j}\right) \prod_{j<m<l}\left(1-2 c^{\dagger}_{m} c_{m}\right)\left(c_{l}^{\dagger}+c_{l}\right)\right\rangle \\
&=\left\langle B_{j} A_{j+1} B_{j+1} \ldots A_{l-1} B_{l-1} A_{l}\right\rangle,
\end{aligned}
\end{equation}
where $A_j=c_j^{\dagger}+c_j$ , $B_j=c_{j}^{\dagger}-c_j$. There are pair contractions for $A_{j}$ and $B_{j}$, i.e.,
\begin{equation}
    Q_{r}=\langle A_j A_l \rangle,
\end{equation}
\begin{equation}
    S_{r}=\langle B_j B_l \rangle,
\end{equation}
\begin{equation}
\begin{aligned}
    G_{r}=&-D_{-r}=\langle B_j A_l \rangle,
\end{aligned}
\end{equation}
where $r=l-j$~\cite{Lee2014}. Given that $G^{xx}_r$ encompasses numerous operators, it is highly essential to represent it using the Pfaffian of an antisymmetric matrix~\cite{wick1950,BE1971}, i.e.,
\begin{equation}
\begin{aligned}
G^{xx}_r=\rm Pf \left|
\begin{matrix}
0 & G_{1} & S_{1} & G_{2} & S_{2} &\cdots & G_{r} \\
 & 0 & D_{0} & Q_{1} & D_{1} & \cdots & Q_{r-1} \\
  &  & 0 & G_{1} & S_{1} & \cdots & G_{r-1} \\
    &  &  & 0 & D_{0} & \cdots & Q_{r-2} \\ 
        &  &  &  & \ddots & \ddots & \vdots \\
         &  &  &  &  & 0 & G_{1} \\
                  &  &  &  &  &  & 0 \\
\end{matrix}
\right|.
\end{aligned}
\end{equation}
In phase transition theory, the spin correlation function $|G^{xx}_r|$ tends to be a fixed non-zero constant in the $\text{AFM}$ phase and decays exponentially to zero in the disordered PM phase. The cross-correlations can be calculated in the same way as the spin correlation function $|G^{xx}_{r}|$. Then, we have
\begin{equation}
\begin{aligned}
G^{xy}_r=i\left\langle B_{j} A_{j+1} B_{j+1} \ldots A_{l-1} B_{l-1} B_{l}\right\rangle,
\\
G^{yx}_r=i\left\langle A_{j} A_{j+1} B_{j+1} \ldots A_{l-1} B_{l-1} A_{l}\right\rangle.
\end{aligned}
\end{equation}

The vector-chiral correlation $|G^{xy}_r|-|G^{yx}_r|$ is a well-defined order parameter to identify the spiral phase for the Ising-Gamma model. In the gapped phases, the vector-chiral correlation $|G^{xy}_r|-|G^{yx}_r|$ tend to be zero due to $|G^{xy}_r|=|G^{yx}_r|$, whereas they remain finite in the spiral phase. An oscillating decline of the vector-chiral correlation $|G^{xy}_r|-|G^{yx}_r|$ as $r^{-1/2}$ with increasing distance $r$ indicates the existence of a quasi-long-range order incommensurate spiral order.

\subsection{Entanglement entropy}
Quantum entanglement is an effective method for characterizing quantum phases and phase transitions~\cite{LSZyu2024,KHN2024}. A widely used measure of entanglement is the entanglement entropy, which is defined by partitioning the system into two subsystems, A and B, and computing the reduced density matrix for subsystem A with size $L$ by tracing out the degrees of freedom of subsystem B:
\begin{equation}
\begin{aligned}
\rho_L = \text{Tr}_B (\left| G_R \right> \left< G_R \right|),
\end{aligned}
\end{equation}
where $\left| G_R \right\rangle$ is the right eigenvector of the non-Hermitian systems in Eq.~\eqref{state}~\cite{MRB2021}.
The entanglement entropy, measuring the entanglement between parts A and B, is then expressed as:
\begin{equation}
\begin{aligned}
S_L&=-\mathrm{Tr}\left[ \rho_L \ln \rho_L \right].
\end{aligned}
\end{equation}

Since the Hamiltonian in Eq.~\eqref{Hami} is quadratic and exactly solvable, the entanglement entropy $S_L$ can be efficiently computed using the correlation matrix method:
\begin{equation}
\begin{aligned}
S_L = - \sum_{j=1}^{L} \left[ v_j \ln v_j + (1 - v_j) \ln (1 - v_j) \right],
\end{aligned}
\end{equation}
where $v_j$ are the eigenvalues of the correlation matrix restricted to the subsystem.

According to conformal field theory (CFT), in (1+1)dimensional critical systems with periodic boundary conditions (PBC), the entanglement entropy obtains a logarithmic correction with
\begin{equation}
\begin{aligned}
S_L&\sim \frac{c}{3}\ln{L},
\end{aligned}
\end{equation}
where $c$ is the central charge. For a broad class of conformally invariant quantum critical points, the central charge $c$ characterizes the universality class of the phase transition in the Hermitian case. Notably, this logarithmic scaling of entanglement entropy also holds in certain gapless critical phases described by CFT.

\begin{figure}[tbhp] \centering
\includegraphics[width=8cm]{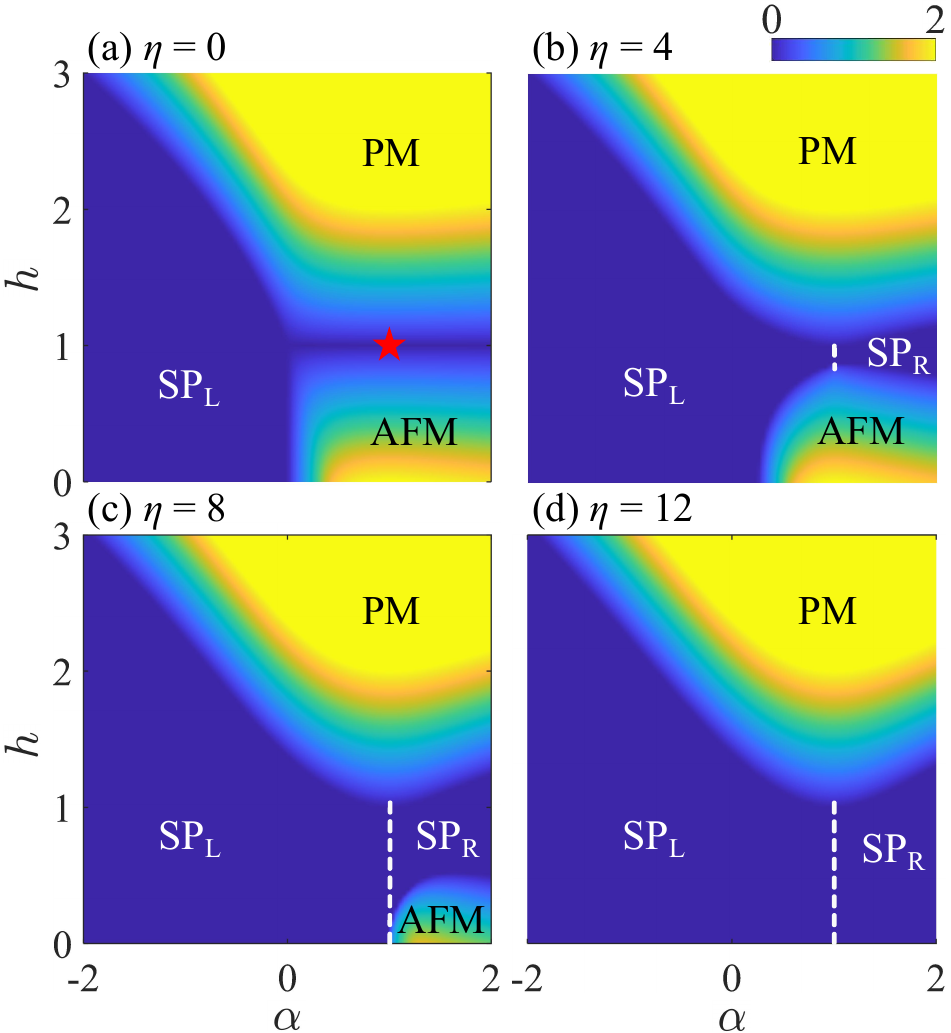}
\caption{The phase diagrams characterized by the real part of the energy gap for $\eta=0$ (a), $\eta=4$ (b), $\eta=8$ (c) and $\eta=12$ (d). The red star denotes the critical point of $\text{AFM}$-PM phase transition when $\alpha=1$ in the Hermitian case. In the non-Hermitian cases, the spiral phase exhibits left-handed chirality to the left of the white line ($\alpha=1$) and right-handed chirality to the right. Throughout, $J=1$, $\Gamma=1$, $N=2000$.}\label{fig1}
\end{figure}

\section{PHASE DIAGRAM}\label{sec:phase}
The schematic phase diagram is provided in Fig.~\ref{fig1}. Let's briefly outline the corresponding phase diagram and summarize the main findings.

Under the condition of $\eta=0$, the model is reduced to the nondissipative case, i.e., the standard Ising-Gamma model. As shown in Fig.~\ref{fig1}(a), there contains three different phases, i.e., $\text{AFM}$, PM and spiral phase. However, the introduction of dissipation will bring about great changes in the phase diagram of the system. When dissipation strength $\eta$ is weak ($\eta=4$), the region of the spiral phase will gradually expand with the division of PM phase and $\text{AFM}$ phase~[see Fig.~\ref{fig1}(b)]. With a further increase in dissipation strength $\eta$ ($\eta=8$), the spiral phase will significantly expand and the $\text{AFM}$ phase region will continuously shrink [see Fig.~\ref{fig1}(c)]. When the dissipation dominates ($\eta=12$), the $\text{AFM}$ phase will disappear completely and only exhibit PM phase and spiral phase in the phase diagram [see Fig.~\ref{fig1}(d)]. Specifically, the spiral phase exhibits left-handed chirality for $\alpha<1$, whereas right-handed chirality is observed for $\alpha>1$, as rigorously proved in Appendix~\ref{sec:Chirality}.

The phase transitions both from the spiral phase to the PM phase and to the $\text{AFM}$ phase are second-order transitions, with the second derivative of the ground-state energy density showing a discontinuity at the critical points.
Specifically, the long-range behaviors of the correlation function along the white lines indicated in the phase diagrams are consistent with the properties of the correlation functions at the critical point in the Hermitian case.

\begin{figure*}[tbhp] \centering
\includegraphics[width=16cm]{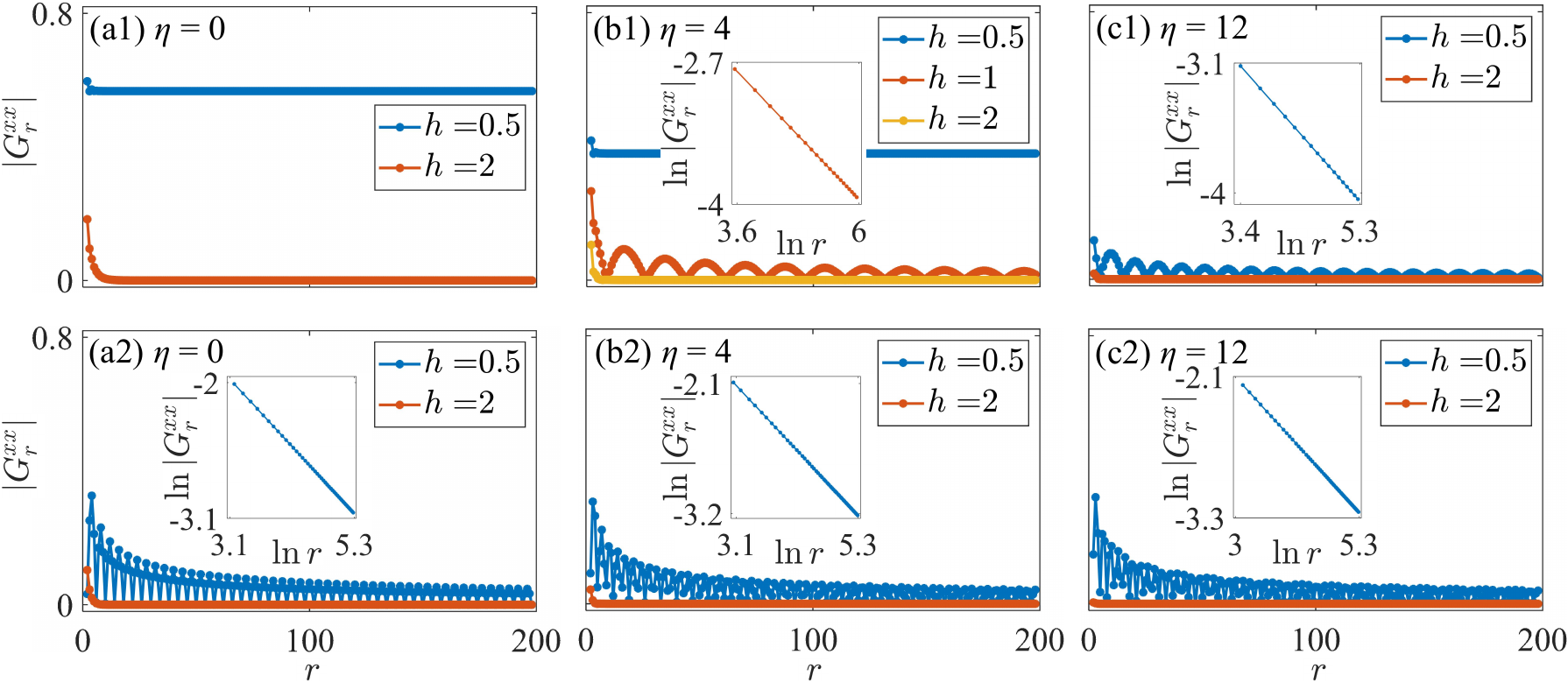}
\caption{The long-range behaviors of spin correlation function $|G^{xx}_r|$ for $\eta=0$ (a1)(a2), $\eta=4$ (b1)(b2), as well as $\eta=12$ (c1)(c2). The parameter $\alpha$ is set to $1.5$ (top row) and $-0.5$ (bottom row). Specifically, the insets show that $|G^{xx}_r|$ presents an oscillating decline as $r^{-1/2}$. Throughout, $J=1$, $\Gamma=1$, $N=2000$.}\label{fig2}
\end{figure*}

\begin{figure*}[tbhp] \centering
\includegraphics[width=16cm]{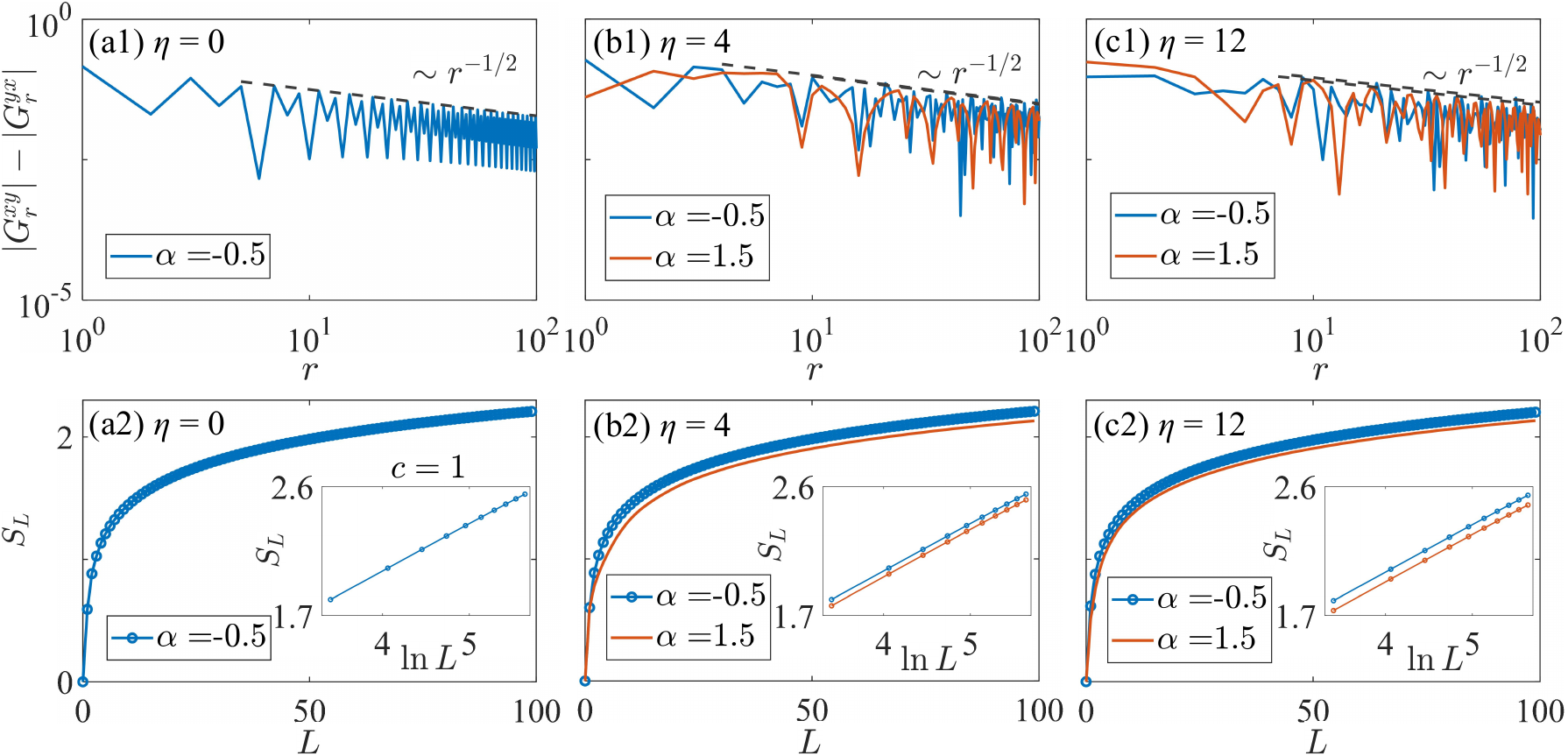}
\caption{The long-range behaviors of the vector-chiral correlation $|G^{xy}_r|-|G^{yx}_r|$ for $\eta=0$ (a1), $\eta=4$ (b1), as well as $\eta=12$ (c1). Specifically, the figures (a1)-(c1) show that $|G^{xy}_r|-|G^{yx}_r|$ presents an oscillating decline as $r^{-1/2}$. The entanglement entropy $S_L$ are plotted against the subsystem size $L$ for $\eta=0$ (a2), $\eta=4$ (b2), as well as $\eta=12$ (c2). The insets in figures (a2)-(c2) highlight the logarithmic scaling behaviors: $S_L \sim \left(1/3\right) \ln L$. Throughout, $J=1$, $\Gamma=1$, $h=1$, $N=2000$.}\label{fig3}
\end{figure*}

\section{The properties of Different phases with dissipation}\label{sec:order}


Now, we explore the possible phases that appear in the phase diagram. Under the condition of $\eta=0$, the model is a standard Ising-Gamma model. By adjusting the parameters $\alpha$, $h$, the model contains three different phases, i.e., antiferromagnetic (AFM), paramagnetic (PM) and spiral phase. Under the condition of $\eta\neq0$, the phase diagrams are greatly different from the Hermitian case [see Fig.~\ref{fig1}].

Now, we present a detailed analysis of the long-distance behaviors of order parameters.

Under the condition of $\eta=0$, the model is a standard Ising-Gamma model. For the case of $\alpha=1.5$, one can find that, when $h=2$, the spin correlation function $|G^{xx}_r|$ tends to be zero, indicating that the system resides in the PM phase [see Fig.~\ref{fig2}(a1)]. When $h=0.5$, $|G^{xx}_r|$ tends to be a constant, which means the corresponding region is the $\text{AFM}$ phase [see Fig.~\ref{fig2}(a1)]. For the case of $\alpha=-0.5$, $|G^{xx}_r|$ presents an oscillating decay as $r^{-1/2}$ when $h=0.5$, suggesting the existence of a quasi-long-range N\'{e}el order in the corresponding region. When $h=2$, $|G^{xx}_r|$ tends to be zero in the long-distance limit, which means the corresponding region is the PM phase under such a circumstance [see Fig.~\ref{fig2}(a2)].

Under the condition of $\eta=4$. For the case of $\alpha=1.5$, the spin correlation function $|G^{xx}_r|$ remains a constant when $h=0.5$, indicating that the system resides in the $\text{AFM}$ phase. In the middle region ($h=1$), the spin correlation function $|G^{xx}_r|$ also exhibits an oscillating decline as $r^{-1/2}$, indicating the corresponding region exhibits a quasi-long-range N\'{e}el order. When $h=2$, $|G^{xx}_r|$ tends to be zero, implying the corresponding region is the PM phase [see Fig.~\ref{fig2}(b1)]. For the case of $\alpha=-0.5$, the spin correlation function $|G^{xx}_r|$ presents an oscillating decline as $r^{-1/2}$ when $h=0.5$, implying the presence of a quasi-long-range N\'{e}el order in this region. When $h=2$, $|G^{xx}_r|$ decays exponentially, confirming that the corresponding region is the PM phase [see Fig.~\ref{fig2}(b2)].

Under the condition of $\eta=12$ ($\eta \gg J$), which means that dissipation dominates, and one can find the $\text{AFM}$ phase vanishes completely in the phase diagram~[see Fig.~\ref{fig1}(d)]. For the case of $\alpha=1.5$, when $h=0.5$, as depicted in Fig.~\ref{fig2}(c1), $|G^{xx}_r|$ presents an oscillating decline as $r^{-1/2}$, also confirming the quasi-long-range N\'{e}el order exhibits in this region. When $h=2$, $|G^{xx}_r|$ remains zero, which means the corresponding region is the PM phase. For the case of $\alpha=-0.5$, when $h=0.5$, the spin correlation function $|G^{xx}_r|$ features an oscillating decline as $r^{-1/2}$, which is consistent with the property of $|G^{xx}_{r}|$ in the spiral phase. When $h=2$, $|G^{xx}_r|$ decays rapidly, suggesting the corresponding region is the PM phase~[see Fig.~\ref{fig2}(c2)].

To indentify the incommensurate spiral order in the system, we calculate the vector-chiral correlation $|G^{xy}_r|-|G^{yx}_r|$. When $\eta=0$, $|G^{xy}_r|-|G^{yx}_r|$ presents an oscillating decline as $r^{-1/2}$ in the spiral phase [see Fig.~\ref{fig3}(a1)], which is consistent with the previous research~\cite{Zhaozhuan2022}. After introducing the dissipation, one can observe that $|G^{xy}_{r}|-|G^{yx}_{r}|$ also shows an oscillating decline as $r^{-1/2}$ [see Figs.~\ref{fig3}(b1)(c1)], suggesting the existence of a quasi-long-range incommensurate spiral order in the corresponding region. Combining with the long-range behaviors of $|G^{xx}_{r}|$, one can confirm that this region is spiral phase.

Additionally, we investigate the scaling behavior of the entanglement entropy. In the spiral phase, it is found that the entanglement entropy follows a logarithmic scaling as $S_L\sim \left(1/3\right) \ln L$, regardless of whether or not dissipation is introduced [see Figs.~\ref{fig3}(a2)-(c2)]. Moreover, by extracting the central charge $c$ from scaling behavior of the entanglement entropy $S_L$, one can diagnose both quantum phases and the phase transition between the PM and $\text{AFM}$ phases in the Ising-Gamma model (see Appendix~\ref{sec:charge} for details).

\section{Phase transitions and critical behaviors}
\label{sec:variation}
After delineating all the quantum phases in the phase diagram, we shift our focus to the more intriguing quantum phase transitions between these phases. In this section, we will investigate phase transitions and critical behaviors. 

\begin{figure}[tbhp] \centering
\includegraphics[width=8cm]{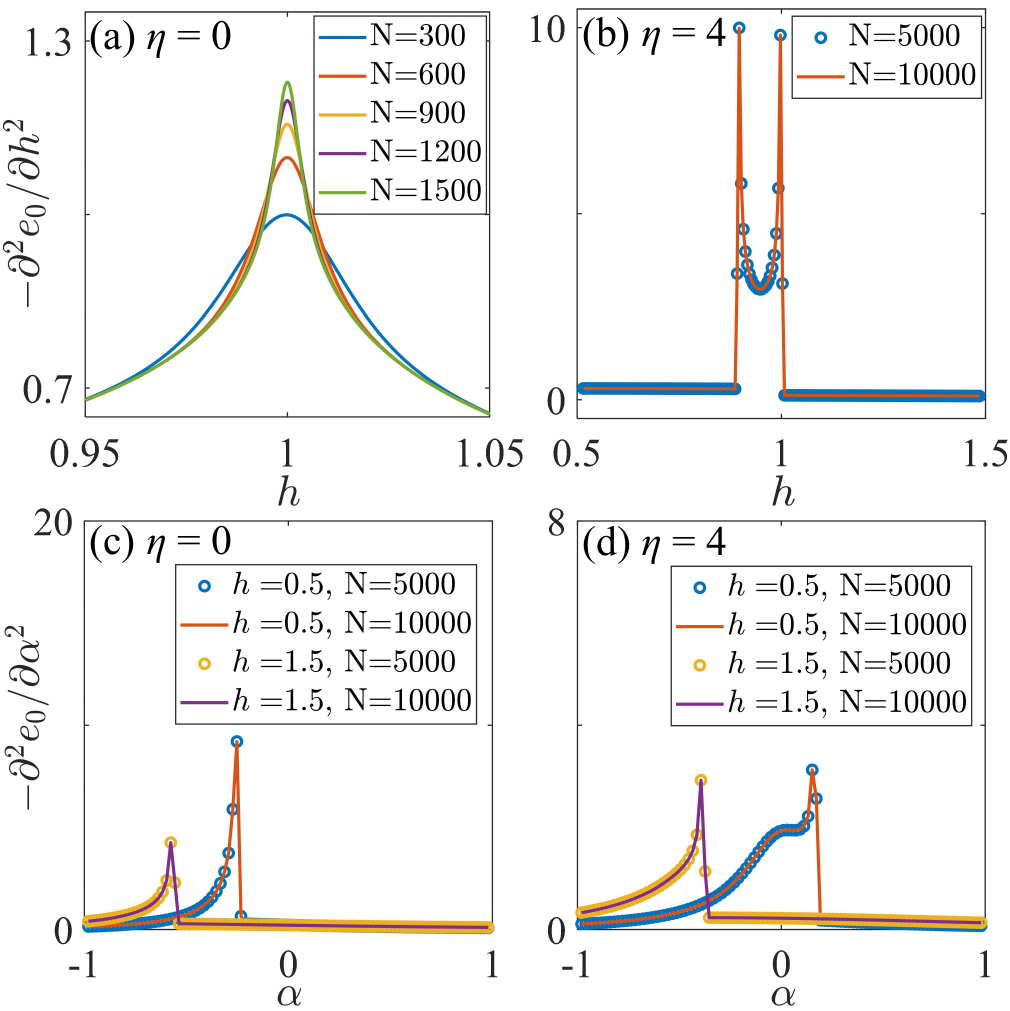}
\caption{The real part of the second derivative of ground-state energy density $-\partial^2 e_0/\partial h^2$ for $\eta=0$ (a), $\eta=4$ (b) in various system sizes with $\alpha=1$. The real part of the second derivative of ground-state energy density $-\partial^2 e_0/\partial \alpha^2$ for $\eta=0$ (c), $\eta=4$ (d) with $h=0.5$ and $h=1.5$ in different system sizes. Throughout, $J=1$, $\Gamma=1$.}\label{fig4}
\end{figure}

When $\eta=0$, $\alpha=1$, we first study the phase transition between $\text{AFM}$ and PM phases. One can observe that the second derivative of the ground-state energy density $-\partial^2e_0/\partial h^2$ presents extreme values around the critical point. With increase of the system sizes $N$, the peaks of $-\partial^2e_0/\partial h^2$ become more pronounced [see Fig.~\ref{fig4}(a)]. By analysing the scaling behaviors of $-\partial^2e_0/\partial h^2$, we obtain the correlation-length exponent $\nu$ of the $\text{AFM}$-PM transition is $\nu\simeq1$ in Appendix~\ref{sec:exponent}.
One can observe that $-\partial^2e_0/\partial\alpha^2$ exhibits the size-independent discontinuity at the critical points of $\text{AFM}$-spiral and PM-spiral phase transitions [see Fig.~\ref{fig4}(c)], which are the same as the results in previous research. Relevant studies also have shown that both $\text{AFM}$-spiral and PM-spiral phase transitions belong to the Lifshitz universality class~\cite{Zhaozhuan2022}.

When $\eta=4$, $\alpha=1$, $-\partial^2e_0/\partial h^2$ exhibits the size-independent discontinuity at two critical points [see Fig.~\ref{fig4}(b)], indicating there emerge new region from the critical point of $\text{AFM}$-PM transitions in the Hermitian case. One can observe that $-\partial^2e_0/\partial\alpha^2$ exhibits the size independent discontinuity at the critical points of $\text{AFM}$-spiral and PM-spiral phase transitions [see Fig.~\ref{fig4}(d)], for $h=0.5$ and $h=1.5$ respectively, which are consistent with the feature of the transition between gapless phase and gapped phases in the Hermitian case.

\begin{figure}[tbhp] \centering
\includegraphics[width=8cm]{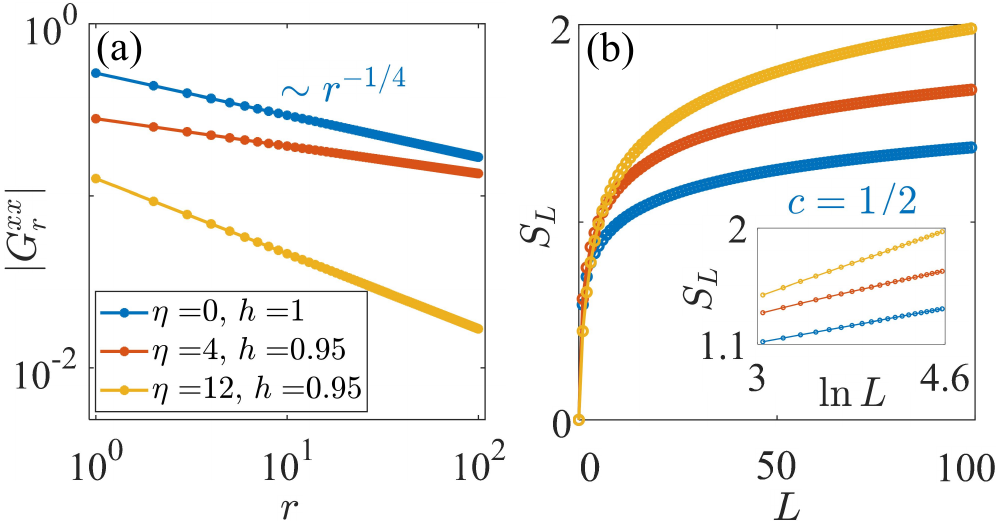}
\caption{(a) The long-range behaviors of spin correlation function $|G^{xx}_r|$ for $\eta=0$, $h=1$ and $\eta=4$, $h=0.95$ as well as $\eta=12$, $h=0.95$. Specifically, $|G^{xx}_r|$ presents a power-law decay as $r^{-1/4}$ when $h=1$ in the Hermitian case. (b) The entanglement entropy $S_L$ are plotted against the subsystem size $L$ for $\eta=0$, $h=1$ and $\eta=4$, $h=0.95$ as well as $\eta=12$, $h=0.95$. Specifically, $S_{L}$ satisfies logarithmic scaling as $(1/6)\ln L$ when $\eta=0$, $h=1$ in the Hermitian case. Throughout, $J=1$, $\Gamma=1$, $\alpha=1$, $N=2000$.}\label{fig5}
\end{figure}

Specifically, we focus on the long-range behaviors of the spin correlation function $|G^{xx}_r|$ and the scaling behaviors of the entanglement entropy in the white line ($\alpha=1$) in the phase diagram [see Fig.~\ref{fig1}]. Under the conditions of $\eta=0$ and external field $h=1$, at the critical point of $\text{AFM}$-PM phase transitions, $|G^{xx}_r|$ exhibits a power-law decay as $r^{-1/4}$, while the entanglement entropy $S_L$ follows a logarithmic scaling. For the case of $\eta=4$, $12$, on the white line, the spin correlation function $|G^{xx}_r|$ exhibits a power-law decay without oscillation, while the entanglement entropy $S_L$ also satisfies the logarithmic scaling [see Fig.~\ref{fig5}]. These behaviors are consistent with those observed at the $\text{AFM}$-PM phase transition point in Hermitian systems.


\section{Summary}
\label{sec:summary}
In summary, we investigate the effect of dissipation on the phase diagram of the Ising-Gamma model. By calculating the energy gap, one can observe that the introduction of complex field can destroy the $\text{AFM}$ and PM phase while simultaneously expanding the spiral phase. By investigating the order parameters and entanglement entropy, we obtain the properties of different phases and the critical behaviors at the points of phase transitions. After introducing the dissipation, in the spiral phase, both the spin correlation function $|G_r^{xx}|$ and the vector-chiral correlation $|G^{xy}_r|-|G^{yx}_r|$ feature oscillating decline as $r^{-1/2}$, and the entanglement entropy satisfies a logarithmic scaling, i.e., $S_{L}\sim\frac{1}{3}\ln L$. These behaviors are consistent with the properties in the Hermitian case. Notably, in contrast to an isolated conservative system, the introduction of dissipation gives rise to the simultaneous emergence of two spiral phases with opposite chiralities. This discovery provides a feasible approach for achieving chirality control of spiral order in artificial quantum systems. Moreover, we demonstrate that the transition between these two spiral phases with opposite chiralities is governed by the relative strength of the off-diagonal Gamma interactions in the Ising–Gamma model. Given that current artificial quantum simulation platforms, such as cold-atom systems, possess the technical capability to realize the Ising-Gamma system and incorporate non-Hermitian dissipation, these theoretical predictions can be directly verified through high-fidelity quantum simulations utilizing cold-atom platforms. This work paves the way for further exploration of emergent quantum phases and phase transitions in open spin systems.


\acknowledgments
This work was supported by the National Key Research and Development Program of China (Grant No. 2022YFA1405300), the Open Fund of Key Laboratory of Atomic and Subatomic Structure and Quantum Control (Ministry of Education) and the Guangdong Provincial Quantum Science Strategic Initiative (Grant No. GDZX2204003).

R.-D. Huang and W.-L. Li contribute equally to this work.

\emph{Data availability}.---The data that support the findings of this article are not publicly available. The data are available from the authors upon reasonable request.

\appendix
\section{Effective Non-Hermitian Hamiltonian}
\label{sec:app0}
In this Appendix, we present the computational details of the non-Hermitian Hamiltonian in Eq.~\eqref{Hami}. The non-Hermitian Hamiltonian in Eq.~\eqref{Hami} can be implemented within the quantum trajectory approach~\cite{PM1998,DZ1992,DJ1992,AJD2014}. Here, we focus on a Markovian open quantum system described by the Lindblad master equation~\cite{AR2011}:
\begin{equation}\label{LME1}
\begin{aligned}
\frac{d}{d t} \hat{\rho}=-i[\hat{H}, \hat{\rho}]+\sum_j\left(\hat{L}_j \hat{\rho} \hat{L}_j^{\dagger}-\frac{1}{2}\left\{\hat{L}_j^{\dagger} \hat{L}_j, \hat{\rho}\right\}\right),
\end{aligned}
\end{equation}
where $\hat{\rho}$ denotes the density operator, $\hat{H}$ is the Hamiltonian that describes the coherent dynamics, and $\hat{L}_j$’s are the jump operators that describe the coupling to the external environment. This master equation can be cast in the form
\begin{equation}\label{LME2}
\begin{aligned}
\frac{d}{d t} \hat{\rho}=-i(\hat{H}_{\rm eff}\hat{\rho}-\hat{\rho}\hat{H}_{\rm eff}^{\dagger})+\sum_j\hat{L}_j \hat{\rho} \hat{L}_j^{\dagger},
\end{aligned}
\end{equation}
with the effective non-Hermitian Hamiltonian
\begin{equation}\label{LME3}
\begin{aligned}
\hat{H}_{\rm eff}=\hat{H}-\frac{i}{2}\sum_j\hat{L}_j^{\dagger}\hat{L}_j.
\end{aligned}
\end{equation}
In the master equation, the term $\hat{L}_j \hat{\rho} \hat{L}_j^{\dagger}$ denotes each quantum trajectory subject to stochastic loss events, and the term $-\frac{i}{2}\sum_j\hat{L}_j^{\dagger}\hat{L}_j$ is the dissipation operator. Under continuous monitoring conditioned on null measurement outcomes (the no-click limit), the dissipative dynamics is governed by the effective non-Hermitian Hamiltonian $\hat{H}_{\rm eff}$. Here we choose the Hamiltonian $\hat{H}$ and the jump operators $\hat{L}_j$ to be 

\begin{equation}\label{LME4}
\begin{aligned}
\hat{H}=&\sum_{j=1}^N J\sigma^x_j\sigma^x_{j+1}+\sum_{j=1}^N\Gamma\left(\sigma^x_j\sigma^y_{j+1}+\alpha\sigma^y_j\sigma^x_{j+1}\right)\\
&+\sum_{j=1}^N h\sigma^z_j,
\end{aligned}
\end{equation}
\begin{equation}\label{LME5}
\begin{aligned}
\hat{L}_j=\sqrt{\eta}\sigma^{-}_{j}.
\end{aligned}
\end{equation}
So the effective Hamiltonian can be written as
\begin{equation}\label{LME6}
\begin{aligned}
\hat{H}_{\rm eff}=\hat{H}-\frac{i\eta}{2}\sum_{j}^{N} \sigma_{j}^{u},
\end{aligned}
\end{equation}
where $\sigma^{u}_{j}=\begin{bmatrix}
1 & 0  \\
0 & 0
\end{bmatrix}$.

\section{Different Chirality of the spiral phase}
\label{sec:Chirality}
\begin{figure}[tbhp] \centering
\includegraphics[width=8cm]{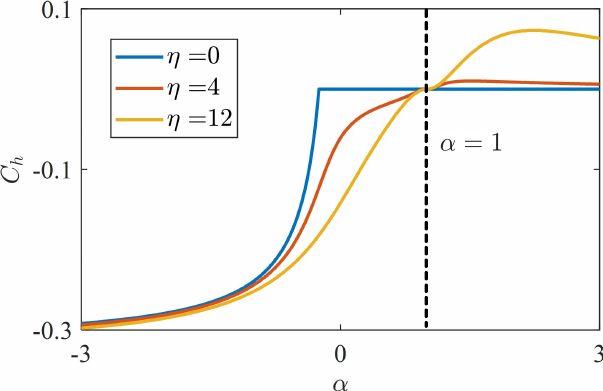}
\caption{Chiral order $C_{h}$ versus $\alpha$ for $\eta=0$, $\eta=4$, as well as $\eta=12$. Throughout, $J=1$, $\Gamma=1$, $h=1$, $N=2000$.}\label{fig6}
\end{figure}
In this Appendix, we investigate the chirality of the spiral phase.
We have calculated the chiral order $C_{h}$ in the $z$ direction as
\begin{equation}\label{CH}
\begin{aligned}
 C_{h}&=G_1^{xy}-G_1^{yx},
\end{aligned}
\end{equation}
which is a well-defined order parameter to identify the chirality of the spiral phase in the Ising-Gamma model. 
The chiral order $C_{h}$ is a positive constant in the right-handed chiral spiral phase, while it becomes negative in the left-handed chiral spiral phase~\cite{RMA2008,KH1988}.

In the Hermitian case ($\eta=0$), when $\alpha<-1/4$, the chiral order $C_h<0$, implying the existence of left-handed chirality in the spiral phase [see Fig.~\ref{fig6}].

After introducing the dissipation ($\eta=4,~12$), when $\alpha<1$, the chiral order $C_h<0$, implying the presence of left-handed chirality [see Fig.~\ref{fig6}]. However, under the condition of $\alpha>1$, we can observe that $C_h>0$, suggesting the existence of right-handed chirality in corresponding region [see Fig.~\ref{fig6}].

The off-diagonal Gamma exchange interactions $\hat{H}_{\Gamma}=\sum_{j=1}^N\Gamma\left(\sigma^x_j\sigma^y_{j+1}+\alpha\sigma^y_j\sigma^x_{j+1}\right)$ can be devided into two parts, i.e., Dzyaloshinskii–Moriya (DM) interaction and the symmetric off-diagonal Gamma interaction,
\begin{equation}
\begin{aligned}
\hat{H}_{\Gamma}&=\sum_{j=1}^NJ^{DM}\left(\sigma^x_j\sigma^y_{j+1}-\sigma^y_j\sigma^x_{j+1}\right)\\
&+\sum_{j=1}^NJ^{SO}\left(\sigma^x_j\sigma^y_{j+1}+\sigma^y_j\sigma^x_{j+1}\right),
\end{aligned}
\end{equation}
where $J^{DM}=\frac{\Gamma(1-\alpha)}{2}$ is the strength of DM interaction, $J^{SO}=\frac{\Gamma(1+\alpha)}{2}$ represents the strength of symmetric off-diagonal Gamma interaction. Obviously, the off-diagonal Gamma interaction reduces to the antisymmetric Dzyaloshinskii–Moriya (DM) interaction and symmetric off-diagonal Gamma interaction for $\alpha=-1$ and $1$, corresponding to $J^{SO}=0$ and $J^{DM}=0$, respectively. The sign of the strength of Dzyaloshinskii–Moriya interaction $J^{DM}$ determines the favoured direction of spins rotation.
For $J^{DM}<0$, a left-handed rotation of spins is promoted, whereas for $J^{DM}>0$, a right-handed rotation is favored~\cite{BPR2025}. When $J^{DM}=0$ ($\alpha=1$), spin rotation is not biased toward any particular handedness, which is consistent with $C_h=0$ in Fig.~\ref{fig6}.

\begin{figure}[tbhp] \centering
\includegraphics[width=8cm]{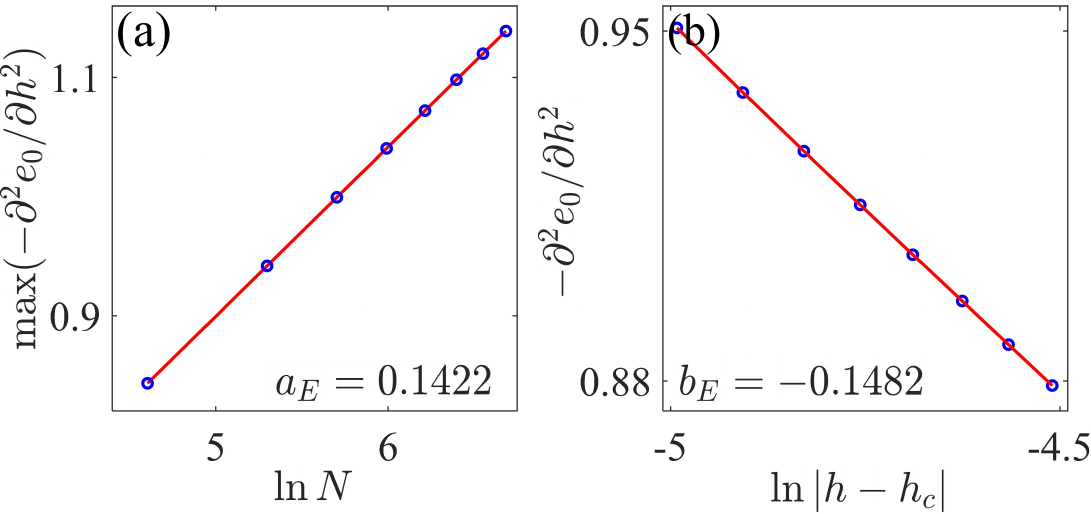}
\caption{(a) The scaling behavior between the maximum value of $-\partial^2 e_0/\partial h^2$ and the system size $N$. (b)  The second derivative of the ground-state energy density $-\partial^2 e_0/\partial h^2$ in the vicinity of the critical point with system size $N=2000$. Throughout, $J=1$, $\Gamma=1$, $\alpha=1$ and $\eta=0$.}\label{fig7}
\end{figure}

\section{The correlation-length critical exponent of $\text{AFM}$-PM transition point in the Hermitian case}
\label{sec:exponent}
In this Appendix, we investigate the correlation-length exponent $\nu$ of $\text{AFM}$-PM transition point in the Hermitian case when $\alpha=1$. A logarithmic singularity across the quantum phase transition between $\text{AFM}$ and PM phase is identified as
\begin{equation}
\begin{aligned}
\max(-\frac{\partial^2 e_0}{\partial h^2}) = a_E \ln N +c_1.
\end{aligned}
\end{equation}
Meanwhile, in the vicinity of the critical point in the thermodynamic limit, one finds
\begin{equation}
\begin{aligned}
-\frac{\partial^2 e_0}{\partial h^2} = b_E \ln |h-h_c| +c_2.
\end{aligned}
\end{equation}
According to the scaling ansatz for logarithmic scaling, the ratio of $|a_E/b_E|$ gives rise to the correlation-length exponent $\nu$. 

As shown in Fig.~\ref{fig7}, the numerical fittings yield $a_{E}=0.1422$ and $b_{E}=-0.1482$, confirming the correlation-length exponent $\nu$ for $\text{AFM}$-PM transitions is $\nu=0.9595\simeq1$.

\section{Central charge as an indicator of quantum phases in the Ising-Gamma model}\label{sec:charge}

\begin{figure}[tbhp] \centering
\includegraphics[width=8cm]{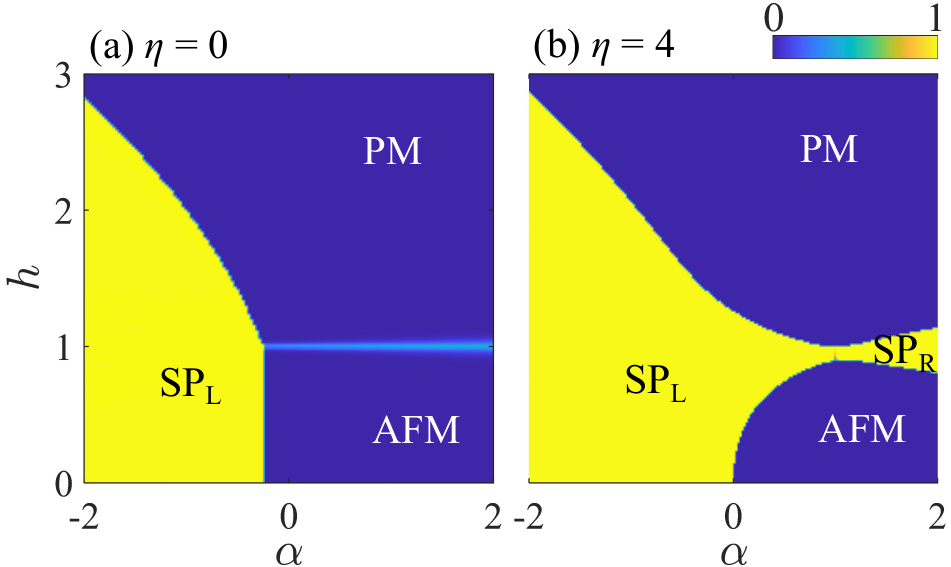}
\caption{The phase diagrams characterized by the central charge $c$ for $\eta=0$ (a), $\eta=4$ (b). Throughout, $J=1$, $\Gamma=1$, $N=2000$.}\label{fig8}
\end{figure}

In this appendix, we demonstrate that the central charge $c$, extracted from scaling behavior of the entanglement entropy $S_L$, can serve as the indicator for identifying quantum phases and the phase transition between the PM and $\text{AFM}$ phases in the Ising-Gamma model.

The entanglement entropy follows the area law in the two gapped phases, while it scales logarithmically in the gapless region \cite{LLL2025}.

By calculating the central charge $c$, we present the phase diagrams of the Ising–Gamma model under both Hermitian and non-Hermitian conditions [see Fig.~\ref{fig8}]. As presented in Fig.~\ref{fig8}, the central charge $c$ has a value of $1$ in the spiral phase, $1/2$ on the critical line between the $\text{AFM}$ and PM phases, and remains $0$ in gapped phases. This result indicates that the central charge $c$ can serve as an effective parameter for identifying the spiral phase and phase transition boundary between the PM and $\text{AFM}$ phases in the Ising-Gamma model under both Hermitian and non-Hermitian conditions.

\bibliography{main}

\end{document}